# The Unpopularity of the Software Tester Role among Software Practitioners: A Case Study

Yadira Lizama[1][0000-0002-0843-8213], Daniel Varona[1][0000-0003-2992-527X], Pradeep Waychal[2][0000-0001-8142-2464] and Luiz Fernando Capretz[3,4][0000-0001-6966-2369]

[1] Cultureplex Laboratories, Western University, London, ON N6A5B9, Canada
{ylizama,dvaronac}@uwo.ca
[2] CRICPE, Western Michigan University, Kalamazoo, MI 49008-5200, USA
pradeep.waychal@gmail.com
[3] Dept. of Electrical & Computer Eng., Western University, London, ON N6A5B9, Canada
[4] Dept. of Computer Science, New York University Abu Dhabi, United Arab Emirates
lcapretz@uwo.ca

**Abstract.** As software systems are becoming more pervasive, they are also becoming more susceptible to failures, resulting in potentially lethal combinations. Software testing is critical to preventing software failures but is, arguably, the least understood part of the software life cycle and the toughest to perform correctly. Adequate research has been carried out in both the process and technology dimensions of testing, but not in the human dimensions. This work attempts to fill in the gap by exploring the human dimension, i.e., trying to understand the motivation/de-motivation of software practitioners to take up and sustain testing careers. One hundred and forty four software practitioners from several Cuban software institutes were surveyed. Individuals were asked the PROs (advantages or motivators) and CONs (disadvantages or de-motivators) of taking up a career in software testing and their chances of doing so. The results of this investigation identified 9 main PROs and 8 main CONs for taking up a testing career showing that the role of tester is perceived as a social role.

## 1 Introduction

Researchers have been investigating practices to improve the work performance of individuals. Several theories were developed and utilized to enlarge the body of knowledge about this theme and to contribute to the improvement of practices. One of the key components which has an impact on the performance and productivity of individuals is the motivation to take up and sustain a job. Nevertheless, software engineering, particularly software testing, still lacks studies on motivation, especially motivation to take up testing careers. Therefore, it is important to focus on phases of the software process, since there are considerable differences in the mindset and skills needed to perform different software tasks.

The success of software projects depends on a precise balance among three pillars: person, process, and technology. Also, the commitment to quality from the team members and a well-organized quality assurance process are essential. In 2015, just a third of software projects were considered a success [1]. Software failures impact the economy [2] [3], cause several technological disasters [4] [5], and have negative social repercussions [6]. The accelerated spread of software in business processes, the increase of complexity and size of systems, and the dependence on third party components are factors that increase potential for failure that impact in software products.

Despite the attention that researchers and practitioners have paid to the process [7] [8] and technology dimensions [9] [10] [18] [19], it is clear that more effort is needed to achieve better results in software testing. It has been pointed out that human and social aspects play a significant role in software testing practices [11] [19] [21]. In an academic setting, attention to human factors in software testing has been preached by Hazzan and Tomayko [12] and Capretz [13].

Aspects of the job that motivate software engineers include problem solving, working to benefit others, and technical challenges, though, the literature on motivation in software engineering appears to present a conflicting and partial picture. Furthermore, surveys of motivation are often aimed at how software engineers feel about the organization, rather than their profession. Although models of motivation in software engineering are reported, they do not account for the changing roles and environmental settings in which software engineers operate.

In a real-world environment, Shah and Harrold [14] and Santos et al. [22] found that software engineers with a positive attitude towards software testing can significantly influence those who have a negative attitude. However, there is no clear understanding of software engineers' jobs in general, what motivates them, how they are motivated, or the outcome and benefits of motivating software engineers.

For a long time, the term motivation was used as a synonym for job satisfaction and to describe several distinct behaviors of software engineers. This satisfaction/motivation disagreement among concepts represented a problem both for academic research and industrial practice, due to the need for the proper management of motivation in software companies, to achieve higher levels of productivity among professionals at work, and motivating software engineers continues to be a challenging task.

Motivational aspects have been studied in the field, including the need to identify with the task in hand, employee participation/involvement, good management, career path, sense of belonging, rewards and incentives, etc. Just like any profession in the world, software engineers also have their own de-motivators, such as the lack of feedback from supervisors, insufficient salary, lack of growth opportunities, etc.

The role of tester does not figure as a favored role among the population of software developers, according to previous study results [15] [19], [20]. Some studies point out the need for reversing people´s perception regarding this role [16] by using career progression and other related mechanisms to reinforce the crucial dimension that a tester brings to the project. In the last decade, individual companies have defined competence profiles for the roles they assign to their projects, as stipulated by the methodology they selected. However, if human aspects are not taken into account in staffing software projects, an important piece of the puzzle for project staffing is overlooked.

We studied the chances of software practitioners taking up software testing careers and their reasons. To that end, we conducted a survey of several Cuban software practitioners; endeavoring to expose actual reactions to the role of tester in the software industry.

## 2     Methodology

In order to assess whether or not participants considered a software testing career, the authors conducted a survey of individuals from a variety of Cuban software institutes. Participants could either be software engineers, software testers, software developers, and/or have an interest in pursuing a career in software testing. The survey asked participants to share the advantages and disadvantages of pursuing a career as a software tester.

The survey asked four questions. The first two questions were open ended questions: 1) What are three PROs (in order of importance) of pursuing a career in software testing; and 2) What are three CONs (in order of importance) of pursuing a career in software testing. The third question asked participants to indicate their intentions of pursuing a career in software testing and were given the option to answer with either "certainly not," "no," "maybe," "yes," and "certainly yes." Participants were also invited to share their reasons behind their responses. Lastly, participants were asked to provide demographic information about themselves. The survey questions are shown in Appendix A.

One hundred and forty-four software developers participated in the survey. 61% of participants were males and 39% were females. The average age of the participants was 31 years old. The average number of testing-related activities among participants was 9 years of experience.

Participants' performance evaluations and current professional roles were not required to participate in the survey. However, once participants consented to and participated in the survey, the authors approached their employers to get access to their performance evaluations as software professionals. These quarterly evaluations assessed professionals (i.e. those who participated in the survey) on a five-point scale and asked the participants' employers to evaluate participants as "Unacceptable," "Acceptable," "Good," "Very Good," and "Excellent"; 85% of participants were assessed as "Good" by the employers, whereas 5% were assessed as "Excellent". The remaining 10% of the assessments were spread across "Acceptable" and "Very Good".

## 3     Results

After a data analysis process, the main results are presented in Table 1. Common statements were combined during the refining of data. We found nine main PROs and eight main CONs in total.

**Table 1.** PROs and CONs descriptions and frequencies.

| No. | FREQUENCY QUANTITY | PERC. | DESCRIPTION |
|---|---|---|---|
| | | | PROs |
| 1 | 71 | 49% | The activities related to testers are simple at first and then complexity is gradually increased; this helps fresh new testers to have a smooth curve of capacitation and specialization. |
| 2 | 27 | 18% | Testing activities provide a full background of project scope, modularization, and integration strategy in a short period of time. |
| 3 | 18 | 12% | The role of tester is a role in which related activities demand lots of creativity from the individual. |
| 4 | 23 | 15% | Test engines and other automated tools give testers accurate support. |
| 5 | 5 | 3% | Tester's responsibilities are spread along all project stages. |
| 6 | 71 | 49% | Tester's activities are quite client oriented. |
| 7 | 52 | 36% | After the analyst role, the tester has more interactions with the project team. |
| 8 | 55 | 38% | A periodical rotation of project team members in the tester role will increase team commitment to product quality. |
| 9 | 42 | 29% | Testing tasks particularities make the software tester focus on details. |
| | | | CONs |
| 1 | 59 | 40% | Other project team members may be upset by a tester´s findings when reviewing their releases |
| | 40 | 27% | |
| 2 | 66 | 45% | Other roles than tester enjoy more acceptance among software engineers. |
| 3 | 72 | 50% | Too many detail-oriented skills are demanded from software testers. |
| 4 | 19 | 13% | Gender-related issues from both parts (males (7) say it is a role for females to perform, and females (12) prefer more technical roles to show their abilities and skills to practitioners of the opposite gender. |
| 5 | 32 | 22% | It is difficult to perform as a software engineer when a person is highly specialized in a particular role. |
| 6 | 39 | 27% | Ability to handle abstraction is needed to have an adequate performance in role tester. |
| 7 | 48 | 33% | The attention of testers has to be divided in two, between engineering artifacts and business process when other roles do not have this division. |
| 8 | 57 | 39% | In some labor markets, the tester wages are less than the average wages of other roles. |

It must be explained that the PROs and CONs were placed, in the table, and ordered reflecting the priority chosen by the respondents. The results related to the PROs and CONs individuals attributed to taking a testing career are presented in the two sections below:

### 3.1 PROs related results

Primarily, the items considered as PROs for taking up a testing career among the surveyed individuals are presented along with their frequencies. The first priority, as can be seen in Table 1, gathers four main trends. The most frequent, with a 36% of respondents, shows the perception that the role of the tester has more interactions with the project team members –better than by the role of analyst (PRO item 7). Followed by the belief of 29% of respondents pointing that testing tasks particularities make software tester focus on details -almost half of individuals in the survey ranked this reason as the second priority (PRO item 9). The remaining two reasons figuring in the first priority, with a 19% and 16% of recurrence respectively, were:

- Testing activities provides a full background of the project scope, modularization, and integration strategy in a short period of time (PRO item 2).
- There are test engines and other automated tools giving testers great technical support (PRO item 4).

In the second priority rank for PROs, the general position was mainly divided into two lines of thought: 49% of the subjects stated that a tester´s activities are client-oriented (PRO item 6) while 47% stated that the particularities of testing tasks makes software tester focus on details (PRO item 9). These PROs were followed by a scarce 3% of individuals who noted that testers' responsibilities are spread within all project stages (PRO item 5).

The third priority contains PROs such as: 49% of responses state that testers start with simpler activities followed by gradual increase in complexity, which helps new testers to have a smooth learning curve (PRO item 1). In addition, 38% suggested that a periodical rotation of project team members in the tester role would increase team commitment to product quality (PRO item 8), and 12% noted the role demands lot of creativity (PRO item 3).

### 3.2 CONs related results

In contrast, the items tagged as CONs for taking up a testing career, respondents gave the most importance to the following reasons:

- Other roles enjoy more acceptance among software engineers (46%) (CON item 2).
- Other project team members may become upset after they receive the tester´s results assessing their work (41%) (CON item 1).

There are gender-related issues in the choice of a testing career (13%) (CON item 4). Eight percent of male subjects in the sample stated that the tester role is a role for females to perform. On the other hand, 21% of female individuals expressed their preference for working in more technical roles such as: programmer or manager, in order to show their competence before a wide male-sexist opinion regarding women in software engineering.

The CONs listed in the second level of importance showed that 50% of the surveyed individual stated that the skills demanded of testers are too detail oriented (CON item 3). 28% of the respondents pointed out the perception that other project team members may become upset facing tester´s findings at the reviewing process (CON item 1); similarly, 41% of individuals supported this same reason within their first priorities regarding cons. The remaining subjects (22%) noted that it is difficult to perform as a software tester when a person is highly specialized in another role, such as analyst or programmer (CON item 5).

Lastly, the third level of importance among CONs was found as follows: 40% of the subjects noted that in the labor market the role of tester is a role for which wages are lower than the average wages for other roles (CON item 8). Another 33% have the idea that the attention of the tester has to cover all engineering artifacts; while other roles only produce a specific type of artifacts (CON item 7). Furthermore, the remaining 27% of respondents believe that the ability to work with abstractions is required to perform adequately in the role of tester (CON item 6).

Regarding the third question, Table 2 shows the responses to the actual chances of respondents taking up a testing career according to their personal preferences. The reason supporting each response for the items chosen are described below. The authors wish to note that similar responses were merged, and duplicates were eliminated to ensure a better understanding and further analysis. A total number of respondents who opted for the response option 'Certainly not' agreed they do not like the role of tester, as did the 25% who chose the response option 'No'. The remaining subjects, who picked the response option 'No', find the role of tester less attractive in comparison to other roles – such as analyst, programmer and designer, in that exact order. Nevertheless, some individuals with the same reason did not specify the role they found more attractive as compared to the tester role. Those individuals who selected the response option 'Maybe', are in total agreement by pointing out that they would perform as testers if no other job offer is available.

**Table 2.** Chances of taking up testing career among respondents.

| THIRD QUESTION DISTRIBUTION | | % |
|---|---|---|
| Certainly not: | 24 | 17% |
| No: | 67 | 46% |
| **Negative subtotal:** | **91** | 63% |
| Maybe: | 22 | 15% |
| Yes: | 23 | 16% |
| Certainly yes: | 8 | 6% |
| **Positive subtotal:** | **53** | 37% |
| **Total of answers:** | **144** | |

## 4   Discussions

According to the results, it can be pointed out that the most recurrent PRO item among respondents was that testing tasks make the tester focus on details. Furthermore, this statement is given most frequently as the second priority CON item. Consequently, the authors believe that focusing on details is a key competence of the role of tester; and at least one core item that software engineers take into account when considering a role.

As it can be seen in previous section, the most common response given as first and second level of importance for both PROs and CONs refers to the testers' interactions with project team members. Therefore, it is accurate to state that the role of tester involves a strong human interaction among software practitioners. The remaining PROs show that subjects identify the role as a way for better approaching a new project when they are newcomers; so, they see the opportunity to improve their soft skills and the demands of creativity as a positive item for their careers.

In addition, it was found some respondents perceive the following two aspects as constructive: the presence of the role of tester through all project stages, and the fact that testing activities provides full access to the project scope, modularization and integration strategy in a short period of time; authors believe that respondents may perceive the role of tester as a professional growth opportunity. Nevertheless, further empirical studies to investigate this aspect of the study needs to be conducted.

During the analysis of the cons, it was noted that the most frequently cited CON item was in regards to team members becoming upset with the tester due to the review of the team members' builds. This could be due to the fact that testers may be accustomed to auditing and criticizing the work of others. Also, the preference for roles other than tester due to their general acceptance constitutes a conclusive statement regarding the unpopularity of the role of tester among respondents.

Some other surprising, if not disturbing, reasons can be found when examining the cons, as it is the case of male-centered issues regarding who is better able to perform the role of tester; males or females. Consequently, a sexist stereotype surrounding individuals performing the tester role was noted; authors finds interesting that the career choices of female subjects are influenced by either male individuals' perceptions, or by their own sexist stereotyped perceptions. The first hypothesis could be the result of women's responses to men believing that the tester role is better performed by women. If added to the unpopularity of the role may re-enforce the devaluations of women; thus women become inclined to do more technical roles in order to prove their equality. On the other hand, the second hypothesis presents the possibility of some sort of feminism that incline women to seek general approval by taking up roles of assumed difficulty in the same way that an attitude of manliness inclines men to do so. Further studies into this area of gendered perceptions around software engineering roles are warranted.

Additionally, it must be highlighted in both PROs and CONs that there were individuals referring to abstraction, creativity and detail-oriented skills when considering to choose the tester role. When comparing percentages, it may be inferred that some of the individuals in the study offered those statements as both PROs and CONs for taking up a testing career in the software industry. The authors agree that to express those statements in either PROs or CONs is a deeply personal point of view, and it can be strongly influenced by the surroundings, the exact period of time when the decision is made, and even the individual's frame of mind. The present study demonstrates that the subjects perceive the role of tester demands: abstraction, creativity, and detail orientation. Further research is needed to clarify whether these adjunctions are exclusive to the role of tester or also conferred upon other roles within software development.

It is surprising that only 40% of subjects, not even a third of the number of participants in the study, refer to the lower wages of the tester role (40 % is more than one third). As if the remuneration was not a factor in the career decision process

When questioned about their chances of taking up a career in software testing, the percentage of negative responses nearly tripled the percentage of positive responses toward the role. An overwhelming quantity of respondents picked the option 'No' as a response. These results concur with prior studies [15] [19], which point to the tester role as one of the less popular roles among others such as project manager, analyst, designer, programmer, and maintenance.

The reasons given to support negative bias towards the tester role are mainly linked to personal preferences, followed by the perception of the role as less attractive than others, as mentioned earlier. On the other hand, supported reasons from those positives response taking up a testing career are related to: similar to personal preferences for most of the subjects in the study, the role's commitment to quality, and the opportunity the role offers to have an overall view of the project in a short time.

The reasons to consider or not a position as tester are directly related to the registered PROs and CONs for question number one in the questionnaire. Meanwhile, reasons supporting the 'Maybe' choice relates to the availability of better job offers, they also may reflect personal preferences and attraction to the role. Furthermore, it is the authors' belief that the tester is a role with more social connotations than technical inclinations, as reflected by the findings of the present investigation.

## 5      Limitations

This study has some intrinsic limitations. Although respondents represent a sample of currently active software practitioners throughout the country, their origin was not recorded. Such data could be used to determine if certain demographic zones are more inclined to certain kind of jobs, specifically testing tasks. Furthermore, the professionals were not asked to state their current role. By doing so, it would be possible to compare the correlation between people's current duties and their career related preferences; primarily those performing as testers at the time the questionnaire was conducted. Nevertheless, these limitations did not compromise the achievement of the study's main goals.

9## 6 Conclusions

The top three PROs reasons for choosing a career as a tester were: (1) testers have more interactions with the project team, after the analyst role; (2) Particularities of testing tasks make the software tester focus on details; (3) Testing activities provide an overview of the project scope, modularization, and integration strategy in a short period of time. With item number 2 being the most frequent PRO item of all within the responses.

The top three CONs reasons for choosing a career as a tester were: (1) Too many detail oriented skills are demanded from software testers; (2) Roles other than tester enjoy more acceptance among software engineers; (3) Other project team members may be upset due to the tester´s findings when their releases are reviewed. With item number 3 being the most frequent CON item of all within the responses.

Given the fact that pro and con more frequently reasons within the responses were: the particularities of testing tasks make software tester to focus on details and how upset a team member could become due to the testers' findings about his/her builds; these are the decisive items for software practitioners when making testing a career of choice.

Among the subjects of the study, the main reasons for taking or not taking up a testing career in the software industry were strongly related to individual preferences and the availability of a job offer involving a more attractive or a better-paid role. Some see the tester role as an opportunity to know quickly what the project entails, and see the benefits of several automated tools that support the tester's performance in their role. In addition, strong human interactions are attributed to the tester role, making it a role with more social connotations than technical implications.

The present study supports prior findings of the unpopularity of the role of tester, positioning the tester role among those less favored by software practitioners. In addition it prompts further research into: (1) why some software engineers considered that testing activities provide a full background of project scope, modularization and integration strategy in a short period of time and the presence of the role of tester through all project stages as a pro; (2) if and at what level, the required abstraction, creativity, and detail orientation are attributed only to the tester role or distributed among the other roles in software engineering; and (3) sexists issues surrounding software engineering roles.

## Appendix A – Survey Questions

We are very grateful to all participants for dedicating their time and attention to our study.

1. What are the three PROs (in the order of importance) for taking up testing career?
    a)
    b)
    c)

2. What are three CONs (in the order of importance) for taking up testing career?
    a)
    b)
    c)

3. What are chances of my taking up testing career?
    Certainly Not    No    Maybe    Yes    Certainly Yes
    Reasons:

4. Gender (optional):

5. GPA (optional):